\documentstyle[aps,prl,preprint,tighten]{revtex}
\begin{document}
\draft
\author{V.N. Bogomolov}
\title{Molecular Crystals and High-Temperature Superconductivity.}
\address{A.F. Ioffe Physicotechnical Institute, Russian Academy of Sciences, St.\\
Petersburg, Russia}
\date{\today}
\maketitle

\begin{abstract}
A simple model of the molecular crystal of $N$ atoms as a statistical
mixture in real space of $NX$ atoms in excited and $N(1-X)$ atoms in
well localized ground state is considered. The phase coherence of the atomic
wave functions is suppose to be absent. A bond energy of crystal is supposed
to be a result of the pair interaction of $NX$ excited atoms. These
molecular type pair excitations do not interact one with another before the
metallization, and do not contribute to the pressure. Nevertheless, the
pressure of such kind of crystals is determined by the interatomic
distances, and by the binding energy of pairs. The possibility of the
insulator-superconductor transition of such a ``gas'' of $NX/2$ pairs,
``dissolved'' among $N(1-X)$ atoms in ground state is discussed. This kind
of transition is supposed to occur in the oxigen $O_2$ , in the sulphur $S$,
and, possibly, in the xenon $Xe$ crystals under pressure. The same kind of
transition is likely to take place in HTSC materials, metal-ammonia and
hydrogen-palladium solutions under normal conditions, due to similarity of
some of their properties with the corresponding ones of molecular crystals.
\end{abstract}


%\begin{multicols}{2}

The problem of finding an adequate description of the type of chemical
bonding, sometimes known as van der Waals bonding in molecular crystals
(MC), is of great practical importance. Bonding of this type plays an
important role in physicochemical processes, such as adsorption,
condensation, catalysis, etc. It can determine some of the properties of
contact between insulators and semiconductors or metals, and the properties
of layered compounds and nanocomposites. Interaction of that type may be
responsible also for the some aspects of the high-temperature
superconductivity (HTSC), regarding the metallization of the MC.

The metallization of condensed gases (MC) is a problem that arose back in
19-th century. Then a supposition was made about the liquid hydrogen to be
metallic \cite{1}. In the first half of the 20-th century the metallic
atomic hydrogen attract an interest as a simplest and superlight alcali
metal analogy. In 1927 a simple criterion of metallization of dielectric
condensates was found \cite{2}. In 1935 the density of atomic hydrogen at
metallization was calculated ($0.6g/cm^3$), and the critical pressure ($\sim
50GPa$ ) was evaluated \cite{3}. Such pressure was out of the experimental
possibilities at that time.

The theory of the van der Waals forces, and the phenomenological theory of
superconductivity were developed by F.London and, perhaps, he was the first
who connected superconductivity with the Bose condensation phenomenon: ``%
{\em ... the degenerate Bose-Einstein gas provides a good example of a
molecular model for such a condensed state in space of momenta, such as
seems need for the superconductive state.}'' \cite{4}. The ``Bose
condensation'' direction in the superconductivity research included
investigations of the systems with the insulator-metal transition from the
insulator side (before the Fermi degeneration). Nevertheless, most of the
studies at that time were carried out on metals (a stable and convenient
objects). The problem of HTSC was absent. Both the high pressure equipment,
and understanding of the insulator-metal physics were absent too. It was the
reason, why the first attempt to undertake the examination of the system
with the insulator-metal transition (metal-ammonia solution) in 1946 was
fail \cite{5}. This system was too complicate. There was only narrow
intermediate interval of the sodium concentrations in ammonia, which lied
between the dielectric and the Fermi metal, in which the anomalous
conductivity was observed. Nevertheless, a small persistent currents was
detected, and upper limit of resistance was about $10^{13} Ohm$ at about $%
200K$. But in 1946 several attempts to reproduce the results \cite{5} was
not successful \cite{6}. Another (and independent of the superconductivity
problem) trend arose in the physics of the metal-insulator transition \cite
{7,8}. As about metals, they was very convenient objects for experiments on
superconductivity, but very intricate for theory. In metal electrons are in
the Fermi-degenerate state, and attempts to transform them into bosons were
not successful \cite{9}. The BCS theory have resolved this problem in 1957.
An attempts to find out the SuperHTSC (``electronic'' mechanism of the
electron pairing) in similar materials were undertaken in 60-th \cite{10,11}
. Very significant progress was attained in the high pressure technique due
to the elaboration of the metallic hydrogen problem \cite{10,12}.
Discovering of the HTSC in 1986 again attract attention to substances with ``%
{\em ...the quasi-metallic character...}'' \cite{5}, which sometimes display
the metal-insulator transition. In 1989 optical properties of metallic xenon
at $140\div 200GPa$ were carefully measured \cite{13}. In 1991 the same
measurements were made on metallic sulphur at $90\div 157GPa$ \cite{14}.
However, in 1997 it was found, that the sulphur under pressure is not a
metal, but a superconductor with critical temperature $T_c\sim 10\div 17K$,
and with fast increase of $T_c$ with pressure \cite{15}. In 1998
insulator-superconductor transition of oxygen $O_2$ ( $90GPa$, $0.6K$) was
observed too \cite{16}. These substances ($S$ and $O_2$) are typical MC, as
well as the condensed xenon $Xe$. The careful inspection of the optics data
\cite{13} allowed to interpret them as a properties of superconductor with
the energy gap $\sim 1.5eV$ and $T_c\sim 5000K$ \cite{17}.

Let us consider some properties of condensed xenon as of a typical MC \cite
{17,18,19}. Only two spectroscopic parameters are quite enough. It is the
ionization potential $E_1=12.13eV$, and the energy difference of the excited
and ground states of the atom $E_2=8.32eV$. The hydrogen-like radii of the
ground and excited states are $r_1=e^2/2E_1=0.593\AA $; $%
r_{21}=e^2/2(E_1-E_2)=1.89\AA $. Radius $r_{21}$ in condensate must be
increased by multiplier about $1.15\div 1.17$ according to well known the
Goldschmidt effect. So, we have $r_{20}=2.2\AA $. This value ($2r_{20}$)
is very close
to the equilibrium interatomic distances in the xenon crystal at the normal
conditions. It means that the condensed xenon may be treated as an ordinary
``chemical'' substance with interaction via the atomic orbitals of excited
state (radius $r_2$), but with their population $X<1$ (``weak excimer
crystal''). Relation $r_{20}/r_1=y^{1/3}$ is about $3.7$, and, according to
the Mott criterion of metallization, the atomic wave functions (radius $r_1$%
) may be treated as a localized ones with the chaotic phases. It allows us
to express the parameter $X$ as $X\sim \exp \left( -E_2/w\right) =\exp
(2)\exp \left[ -\left( y^{1/3}+y^{-1/3}\right) \right] $, where $%
w=e^2/2\left( r_2-r_1\right) $, --- the mean energy of the chaotic
interatomic interaction of the ground state electrons. At normal conditions $%
X_0=0.14$ for xenon. In reality it means, that there are $NX$ atoms in
excited state (the excimer analogy of $Cs$) and $N(1-X)$ atoms in the
ground state among $N$ atoms of condensate at every moment. It leads to the
conclusion, that the excited atoms exist by pairs, to take part in bonding
of crystal. These pairs are hydrogen-like virtual molecules $(Xe)_2$ of
excimer xenon atoms. Their bond energy is $Q_{20}=0.28e^2/2r_{20}\sim 0.9eV$
as for the covalent type molecules. The mean bond energy of condensed xenon
is $q_0=X_0Q_{20}\sim 0.13eV$. The compressibility
$k\sim 30\,10^{-6}atm^{-1}$
of condensed xenon is also close to the calculated value: $%
k_c=0.16r_{20}^4/X_0\sim 27\,10^{-6}atm^{-1}$. One can obtain the equation
of state (EOS) :
\begin{equation}
P(y)=P_0E_1^4\int (y-2)^{-7/3}\exp \left\{ -\left[
(y-2)^{1/3}+(y-2)^{-1/3}\right] \right\} \,dy\,.  \label{1}
\end{equation}
The transition from $y$ to $y-2$ is a result of subtraction the ``dead''
volume $\sim r_1^3$ from the volume $\sim r_2^3$, and of the bond energy $Q$
from $E_1$. This EOS satisfactory corresponds to the experimental data for $%
Xe$ \cite{20}. This model does not contradict with the results of the
traditional description of MC, utilizing the superposition of the ground and
the excited state wave functions,
and seems to be more adequate for the ``Mott situation''.
The same approach is applicable for the
sulphur $S$, and for the oxygen $O_2$ condensates (excimer molecules are $%
(S)_2$ and $(O_2)_2$ ) \cite{19}.

At metallization the molar volume of xenon is $10.27cm^3/mol$ ($34.7cm^3/mol$
at normal conditions). It means that $r_m=r_{2{\rm si}}=1.47\AA $, $y_{{\rm %
si}}^{1/3}=2.47$, which is about the Mott criterion of metallization, and
corresponds to the band structure appearance. Parameter $X_{{\rm si}}\sim
0.42$. For the pair excitations (the ``Frenkel biexcitons''), $X_{2{\rm si}%
}\sim 0.21$. It is about the percolation threshold for the xenon ``site''
lattice.

For the metallic state the bond energy of molecules $(Xe)_2$ is $Q_{2{\rm si}%
}=0.28e^2/2r_{2{\rm si}}\sim 1.37eV$. It is close to the energy gap ($\sim
1.5eV$) of hypothetic superconductivity state, evaluated for
``metallic''xenon in \cite{17} from the optical data \cite{13}.
Metallization of xenon likely is an insulator-superconductor transition too.

Near the transition (from the insulator side) there are ``gas'' of
noninteracting hydrogen-like $(Xe)_2$ molecules, distributed over the
lattice sites and not contributing to the pressure of crystal (Fig.\ref{fig1}%
, insertion). Their bond energy, and elastic properties of crystal are
related with the existence of electron pairs, having zero momentum and spin
(bosons). External pressure changes only bond energy of these electron pairs
(and their concentration) by changing the interatomic distances. The
situation is similar to dependence of the Cooper pair energy on the dynamic
properties of lattice. The possibility of the stable situation below the
equilibrium distance $r_{20}$ under pressure is shown in Fig.\ref{fig1}
(curve $E_1$ and $E_3$). Curve $E_2$ corresponds to the ``unstable lattice''
of $ns^1$ or $ns^2$ atoms at $r_2>r_{10}$, --- the equilibrium
distance. At $r_2>r_{2{\rm si}}$ we have the insulator state in
both cases. Below $r_{2{\rm si}}$ the tunneling processes, and possible
superconductivity (for lattice of metal atoms, and for clusters of $(Xe)_2$%
, $(S)_2$ and $(O_2)_2$ molecules) becomes significant.
In this region MC looks like
a granular superconductors with the Josephson tunneling (Fig.\ref{fig1},
insertion). The intermolecular interaction energy of the excimer molecules,
for example of $(Xe)_2$, is about zero up to the Bose or to the Fermi
degeneracy point. The Bose degeneracy temperature for the xenon crystal is $%
T_B=0.084h^2(NX_{2{\rm si}})^{2/3}/2m_e\sim 5500K$. It is close to values
obtained above. The bond energy per one $(Xe)_2$ boson is $Q_B\sim 1/r_2$
(Fig.\ref{fig1}). The mean energy $Q_F$ per one $(Xe)$ fermion
for the ``Fermi gas'' approximation is proportional to $1/R^2\sim
(NX)^{2/3}\sim 2Q_Br_2^{-1}\exp \left( -2r_2/3r_1\right) $ (Fig.\ref{fig1}
). Bosons seem to be more preferable than fermions for the
interval $r_{2{\rm ms}}-r_{2{\rm si}}$. The situation is quite opposite at $%
r_2<r_{2{\rm ms}}$. Below $r_{2{\rm ms}}$ evolution of the both systems ($E_1
$ and $E_2$) is likely demonstrates Fermi degeneracy.
It should be noted, that the palladium is the only metal with interatomic
distances, corresponding to $r_2$ ($\sim 1.3\AA$), but not to $r_1$
($0.56\AA$) \cite{21a}.
Such kind of a
insulator-superconductor transition is likely occurs for the oxygen ($0.6K$%
), sulphur ($17K$) and, perhaps, xenon ($\sim 5000K$). Most atoms in sulphur
and molecules in oxygen (namely, $N(1-X)$) are in ground state, they
possess magnetic momenta (especially $O_2$), and play the role of magnetic
impurities \cite{19}. For the sulphur magnetic susceptibility is $\sim
1/(w+kT)\ll 1/kT$, and may be neglected, but not for the electronic
processes. The critical temperature $T_c$ is known to be strongly dependent
on concentration of the magnetic impurities \cite{21}. The situation for $Xe$
is much better, because the xenon atoms in ground state are diamagnetic.

In the HTSC oxide type materials the ion $O^{2-}$ is probably in an excited $%
3s^2$ state \cite{19}. Together with $ns^{2+}$ ions this system resembles MC
at the threshold of metallization (superconductivity).

When the insulator-insulator contact is changed by the insulator-metal one
(see Fig.\ref{fig2}), several effects arise:

\begin{enumerate}
\item  Here, too, the ``vacuum'' gap ($r_2-r_1$) between the insulator and
metal appears. It was clearly observed for the first time
in capillaries of atomic diameters in zeolites
\cite{22}.

\item  Atoms or molecules of insulators in contact with atoms of metal
become ``virtual excimers'' ($X>0$), and receive strong chemical activity
(catalytic action).

\item  Parameter $X$ increases, because the perturbation energy $%
w=e^2/2(r_2-r_1)$ is changed by $w_m=e^2/(r_2-r_1)$. For $Xe$ decrease of $%
r_2$ by factor 1.5 corresponds to pressure about $150GPa$. At the surface of
the metal, the insulator may be transformed into the superconductor as if
under high pressure. The magnetic and ``proximity'' effects should decrease $%
T_c$. Possible examples of such a situation are the sodium (metal) atoms in
ammonia (MC and the molecular excimer type pairs $(NH_3)_2$), and the
hydrogen (MC and the molecular excimer type pairs $(H)_2$) atoms in
palladium (metal). The volume of the solution increases with the appearance
of the ``interface gap'', and of the volume difference $\symbol{126}%
(r_2^3-r_1^3)$ in both cases.
\end{enumerate}

The MC has a very significant difference from the covalent crystals, and
from the metals with well defined band structures. The Mott criterion for
the MC corresponds to insulator state, and to well localized atomic wave
functions with chaotic phases. It is a reason to utilize the mixture of
atoms in excited and ground states in real space, instead of the
superposition of the atomic wave functions and the band theory description
of MC. The electronic properties (and the bond energy) of the weakly bonded
MC depends on the fluctuations of the electron-electron interaction energy $%
w\sim e^2/2(r_2-r_1)$, but not on the dynamic properties of the lattice and
on the thermal energy. When this part of the electronic subsystem of atoms
becomes correlated, and the band structure appears, the phase transition of
the MC into the covalent crystal happens. The mean fluctuation energy $w$
changed by the constant $w_0\ll w$.

The situation looks like if in the ordinary semiconductor one changes the
dynamic properties of the lattice by that of the interacting electrons
subsystem of atoms in ground state. The mean energy $kT$ must be changed by
the mean electron-electron interaction energy $w(r)$. The concentration $X$
of electrons within the excited state orbitals (in the conducting band)
depends on the interatomic distance (pressure) but not on the heat.
Molecular electron pairs in excited state (bosons) resembles small
bipolarons in ordinary lattice. Phonons must be changed by the
``homoplasmons'' (fluctuations of the electron density). A simple model of
such a statistical
mixture of atoms in real space with random field interaction allow us
to made some conclusions on the possibility
of the boson--like excitations existance, and
of the electronic mechanism of
the superconductivity of the MC and of the HTSC materials.

It is seen from Fig.\ref{fig1} (curve $E_3$), that superconductivity is
probable for the compressed MC at $r<r_{20}$, or for the metals at $r>r_{10}$
(e.g. disordered systems,
insulator-metal solutions,
or compounds). Some hypothetical metal-insulator
``nanocomposite'' with regular lattice of heteroclusters (for example, $(%
{\rm Metal})Xe_3$, $({\rm Metal})Xe_{12}$, etc.) appeared to be interesting
materials to search for the high-temperature superconductivity.

\begin{figure}[tbp]
\caption{Schematic representations of the bond energy dependences
on the half of the interatomic distance $r_2$ for the
molecular ($E_1$), for the covalent ($E_2$) type crystals,
and for the superconductor ($E_3$). Also the
mean bond energy of the molecular type pairs $Q_B$ for the MC, and
the mean fermion energy $Q_F$ for the metal are shown.
Radii $r_1$ (covalent one) and $r_2$ (van der Waals one)
corresponds to ground and excited state
orbitals of atom. Intermolecular distances for the MC are equal to $R$
(insertion).}
\label{fig1}
\end{figure}

\begin{figure}[tbp]
\caption{Schematic representation of the insulator--insulator (A-B) and
insulator--metal (A-M; B-M) contacts. $E_{AB}=e^2/2(r_2-r_1)$,
$E_{AM}\sim E_{BM}=e^2/(r_2-r_1)=2E_{AB}$}
\label{fig2}
\end{figure}

%\end{multicols}

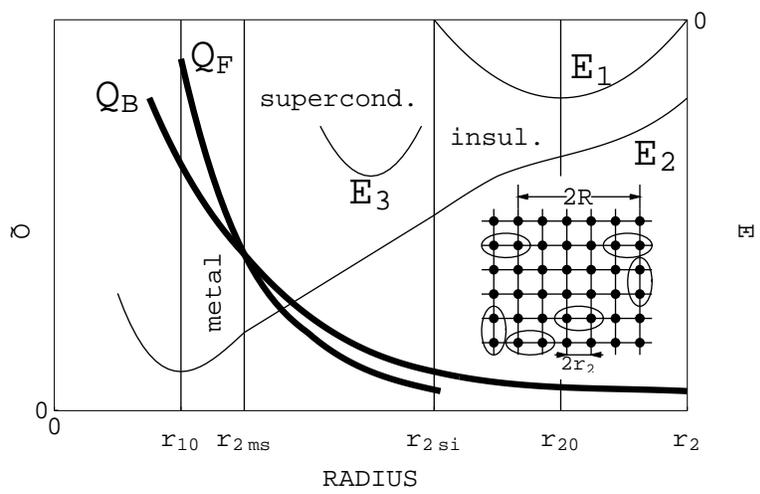
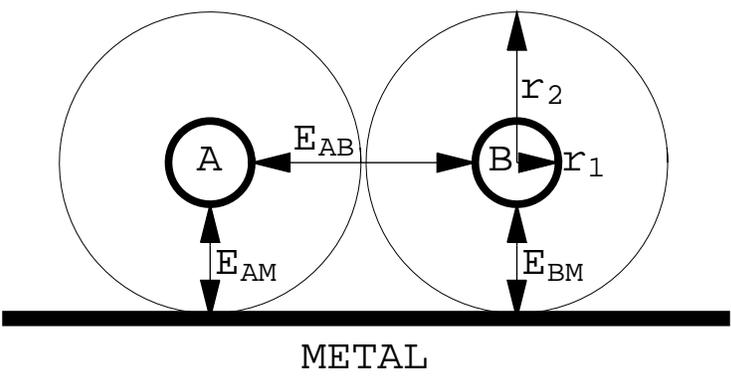
\end{document}